\documentclass[a4paper]{article}

\usepackage{INTERSPEECH2021}
\usepackage{enumitem}

\DeclareMathOperator{\sinc}{sinc}
\DeclareMathOperator{\rect}{rect}

\title{End-to-end Keyword Spotting using Neural Architecture Search and Quantization}
\name{David Peter, Wolfgang Roth, Franz Pernkopf}
\address{
  Signal Processing and Speech Communication Laboratory\\
  Graz University of Technology\\
  Graz, Austria}
\email{david.peter@student.tugraz.at, roth@tugraz.at, pernkopf@tugraz.at}

\begin{document}

\maketitle
\begin{abstract}
This paper introduces neural architecture search (NAS) for the automatic discovery of end-to-end keyword spotting (KWS) models in limited resource environments. We employ a differentiable NAS approach to optimize the structure of convolutional neural networks (CNNs) operating on raw audio waveforms. After a suitable KWS model is found with NAS, we conduct quantization of weights and activations to reduce the memory footprint. We conduct extensive experiments on the Google speech commands dataset. In particular, we compare our end-to-end approach to mel-frequency cepstral coefficient (MFCC) based systems. For quantization, we compare fixed bit-width quantization and trained bit-width quantization. Using NAS only, we were able to obtain a highly efficient model with an accuracy of 95.55\% using 75.7k parameters and 13.6M operations. Using trained bit-width quantization, the same model achieves a test accuracy of 93.76\% while using on average only $2.91$ bits per activation and $2.51$ bits per weight.

\end{abstract}

\section{Introduction}
Automatic speech recognition (ASR) is becoming increasingly important for user interaction with everyday consumer devices. ASR systems are typically complex and computation-intensive, i.e. running ASR in always-on mode results in a steady high energy consumption. This is especially problematic for mobile devices whose batteries are drained quickly when running ASR permanently. 

A common solution is to run a low-cost keyword spotting (KWS) system that is listening permanently only for a limited set of prespecified keywords. Upon detection of such a keyword, a full ASR system is triggered which then listens for a rich set of user commands. The requirements of a KWS system are: (i) The system should be resource-efficient to mitigate the aforementioned energy problem, (ii) it should run in real-time and, (iii) it should be accurate to maintain a high user-experience.

In KWS, deep neural networks (DNNs) have become the state-of-the-art. In \cite{Zhang2017} several models from the literature \cite{Chen2014,Sainath2015,Arik2017,Sun2016} are evaluated on the Google speech commands dataset \cite{Warden2018}. They compare their models in terms of accuracy, memory requirements and number of operations (i.e. the number of multiplications and additions)  per forward pass. To allow for easy deployment on microcontrollers they train their models using 32 bit float numbers and quantize the weights after training to 8 bit fixed-point numbers. They argue that fixed point numbers have been shown to suffice to run DNNs with minimal loss in accuracy \cite{Suda2016,Qiu2016,Lai2017}.

There are many aspects to consider when designing resource-efficient DNNs for KWS. In this paper, we will focus specifically on the following important aspects: The DNN architecture, comparison between spectral MFCC feature and raw audio processing as well as quantization of weights and activations.

Recently, neural architecture search (NAS) became a popular technique to automate the design of DNN architectures. A NAS algorithm searches for the best performing DNN architecture within a given search space using appropriate search heuristics. Popular NAS approaches use concepts such as reinforcement learning \cite{Pham2018}, gradient based methods \cite{Liu2019} or evolutionary methods \cite{Liu2018} for exploring the search space. In the context of resource-efficient DNNs, NAS techniques have also been used to find DNN architectures that are specifically tailored to the underlying hardware \cite{Cai2019,Tan2019} by, for instance, additionally minimizing memory requirements, number of operations, or latency of the resulting model. Therefore, NAS techniques are well-suited for finding DNNs that run on mobile phones or embedded devices. In the context of KWS, NAS has been used successfully by \cite{Peter2020,Mo2020} to find efficient and small DNN architectures.

KWS with DNNs is typically performed on hand-crafted speech features such as MFCCs that are extracted from raw audio waveforms. Extracting MFCCs involves computing the Fourier transform. However, performing the Fourier transform is computationally expensive and might exceed the capabilities of resource-constrained devices. Therefore, Ibrahim et al. \cite{Ibrahim2019} proposed to use simpler speech features derived in the time-domain. The features, referred to as Multi-Frame Shifted Time Similarity (MFSTS), are obtained by computing constrained lag autocorrelations on overlapping speech frames to form a 2D map. A temporal convolutional neural network (TCN) \cite{Chiu2018} is then used to classify keywords on MFSTSs. 

However, hand-crafted features such as MFSTSs and MFCCs may not be optimal for KWS. Therefore recent works have proposed to directly feed the DNNs with raw audio waveforms. In \cite{Ravanelli2018}, a CNN for speaker recognition is proposed that encourages to learn parametrized sinc functions as kernels in the first layer. This layer is referred to as SincConv layer. During training, a low and high cutoff frequency per kernel is determined. Therefore, a custom filter bank is derived by training the SincConv layer that is specifically tailored to the desired application. SincConvs have also been recently applied to KWS tasks \cite{Mittermaier2020}.

Quantization-aware training uses the straight-through estimator (STE) \cite{Hinton2012,Bengio2013} to approximate the gradient of piecewise constant quantizers by the non-zero gradient of some other function. The STE has been used for example in the training of binarized neural networks (BNNs) where the resolution of both weights and activations is reduced to binary values $\{-1,1\}$. Another method for quantizing DNNs involves a Bayesian approach to learn weight distributions over discrete weights \cite{Roth2019}. Recently, the STE has been used to also learn the weight and activation bit-widths (i.e. number of bits) during training \cite{Uhlich2020}. We will refer to this type of training as\emph{ trained bit-width quantization}.

In this paper, we propose NAS for the automatic discovery of small and resource-efficient end-to-end models for KWS. We utilize the techniques from ProxylessNAS \cite{Cai2019} to find suitable end-to-end KWS models. During NAS, we establish a tradeoff between the model accuracy and the number of operations. We compare our end-to-end KWS models obtained by NAS to MFCC based systems. Once the efficient full-precision end-to-end KWS model has been found, we compare two weight and activation quantization methods. Both methods perform quantization-aware training from scratch on the full-precision end-to-end KWS model. In the first method, the weight and activation bit-widths are fixed during training. In the second method, trained bit-width quantization is performed.

\noindent
Our contributions are the following:
\begin{itemize}
	\item We apply NAS to obtain efficient end-to-end KWS models operating on raw audio waveforms instead of hand crafted features. Recent works such as \cite{Mo2020,Zhang2020,Peter2020} rely on hand crafted features such as MFCCs when performing NAS for KWS.
	\item We perform a thorough comparison between KWS on raw audio waveforms and KWS on MFCCs in terms of accuracy, number of operations and number of model parameters.
	\item We compare two quantization methods for weight and activation quantization. In particular, we perform fixed bit-width and trained bit-width quantization on end-to-end KWS models to further reduce the memory footprint. 
\end{itemize}

The outline of the paper is as follows: In Section~\ref{sec:methods} we present our NAS configuration, the feature extraction using SincConvs and the weight quantization methods utilized in this paper. The experimental setup is shown in Section~\ref{sec:experimental_setup}. In Section~\ref{sec:experiments}, we discuss the results of our experiments. Finally, Section~\ref{sec:conclusions} provides the conclusion.

\section{Methods}
\label{sec:methods}

\subsection{Neural architecture search}
Our goal is to find well performing architectures for different computing regimes. To achieve this, we use multi-objective ProxylessNAS \cite{Cai2019} to discover DNNs optimized for accuracy and number of operations. Note that this implicitly optimizes for the model size as well.

ProxylessNAS constructs an overparameterized model with multiple parallel candidate operations per layer as the base model. The overparameterized model is trained together with a set of architecture parameters specifying probabilities over the candidate operations. Once training has finished, for each layer the most probable candidate operation is selected.

Table~\ref{model_structure} shows the overparameterized model used in this paper. It consists of five stages with two input stages (i) (ii), two intermediate stages (iii) (iv) and one output stage (v). Stages (i), (ii) and (v) are fixed whereas stages (iii) an (iv) are optimized using NAS. We use mobile inverted bottleneck convolutions (MBCs) \cite{Sandler2018} as our main building blocks in stages (iii) and (iv).

MBCs have two learnable parameters, the expansion rate $e$ and the size $k$ of the quadratic $k$$\times$$k$ convolution kernel. MBCs consist of three separate convolutions, one 1$\times$1 convolution followed by a depthwise-separable 3$\times$3 convolution followed again by a 1$\times$1 convolution. The first two convolutions apply batch normalization and ReLU activation functions. The third convolution only applies batch normalization. The first and third 1$\times$1 convolution change the number of feature maps by the expansion rate factor of $e$ and $1/e$ respectively. Stride (as stated in Table~\ref{model_structure}) is only applied to the first convolution of each stage.

During NAS, we allow MBCs with expansion rates $e \in \{1,2,3,4,5,6\}$ and kernel sizes $k \in \{3,5,7\}$ for selection. We also include the zero operation which effectively results in an identity layer [13]. For blocks where the input feature map size is equal to the output feature map size we include skip connections.

\begin{table}[t!]
	\caption{NAS model used for KWS. $K$ denotes the kernel size, $S$ the stride, $C$ the number of channels and $L$ the number of layers per stage. Stages (i) and (ii) and (v) are fixed. For stage (iii) and (iv), the parameters $e$ (expansion rate), $k$ (kernel size) and whether an identity layer is selected or not is optimized using NAS.}
	\begin{center}
		\begin{tabular}{cccccc}
			\toprule
			\textbf{Stage} & \textbf{Type} & $\textbf{K}$ & $\textbf{S}$ & $\textbf{C}$ & $\textbf{L}$ \\
			\midrule
			(i) & SincConv & 400 & 160 & 1 & 1 \\

			(ii) & Conv & 3x3 & 2, 2 & 10 & 1 \\

			(iii) & MBC[$e$] / Identity & [$k$]$\times$[$k$] & 2, 2 & 20& 3 \\

			(iv) & MBC[$e$] / Identity & [$k$]$\times$[$k$] & 2, 2 & 40 & 3 \\

			(v) & Conv & 1$\times$1 & 1, 1 & 80 & 1 \\
			& Global Avg. Pooling & - & - & - & 1\\
			& Fully connected & - & - & - & 1\\
			\bottomrule
		\end{tabular}
	\end{center}
	\label{model_structure}
\end{table}

\subsection{Feature extraction using SincConvs}
SincNet \cite{Ravanelli2018} uses parametrized sinc functions as filters in the first layer. This layer performs a convolution of the raw audio input $x$ with an arbitrary number of parametrized sinc functions called SincNet filters. For a single SincNet filter, the output $y$ given the filter $g[n, \theta]$ parametrized by $\theta$ is simply
\begin{equation}
	y[n] = x[n] * g[n, \theta].
\end{equation}
In SincNet, the choice for the filter function $g$ is
\begin{equation}
	g[n, f_1, f_2] = 2f_2 \sinc(2\pi f_2 n) - 2f_1 \sinc(2\pi f_1 n),
\end{equation}
where the sinc function is defined as $\sinc(x) = \sin(x) / x$ and $\theta = (f_1, f_2)$. This choice of $g$ can be seen as a bandpass filter in the frequency domain with $f_1$ and $f_2$ being the low and high cut-off frequencies of the bandpass filter. The magnitude of the bandpass filter in the frequency domain is therefore
\begin{equation}
	G[f, f_1, f_2] = \rect\left(\frac{f}{2 f_2}\right) - \rect\left(\frac{f}{2 f_1}\right)
\end{equation}
where $\rect(f)$ is the rectangular function defined as
\begin{equation}
	\rect(f) = \begin{cases} 1 &\mbox{if } |f| \leq \frac{1}{2} \\
		0 & \mbox{if } |f| > \frac{1}{2} \end{cases}.
\end{equation}
SincConv filters have a much smaller memory footprint than 1D-Convs where the filter kernel is fully learnable. To derive a single filter $g[n, f_1, f_2]$, only two parameters, the lower cutoff frequency $f_1$ and the upper cutoff frequency $f_2$ are needed, whereas for convolutions with arbitrary filters, the number of parameters to store is equal to the length of the filter.

During runtime however, SincConvs and 1D-Convs of similar length need the same amount of memory to store the filter kernels. SincConv filter kernels are precomputed once before runtime. However, the computational cost of computing the SincConv filter kernels can typically be neglected.

\subsection{Weight and activation quantization}
For the architecture discovered by NAS, we compare two quantization methods for quantization of weights and activations, namely fixed bit-width quantization and trained bit-width quantization.

We perform quantization-aware training using the quantization framework brevitas \cite{brevitas}. We quantize weights and activations of all layers, except for the input layer.

In quantization-aware training, quantized tensors are obtained from real-valued auxiliary tensors by applying a quantization function $Q$. During backpropagation, the gradients of the auxiliary tensors are obtained using the STE to estimate the gradient of $Q$. The quantized tensors are typically integer numbers encoded to $k$ bits. A factor $\alpha$, called the dynamic range \cite{Uhlich2019,Jain2020,Esser2020}, is used to map the integer numbers of the quantized tensor to real-valued numbers. The scaling factor typically increases the performance quite substantially.

For quantized weights, we select $\alpha$ to be the maximal absolute value of the auxiliary weight tensor. For quantized activations, we select $\alpha$ differently depending on the bit width. Binary (i.e. 1 bit) quantization of activations is performed using a constant scaling factor of $\alpha=1$. When the activation bit-width is larger or equal to 2 bits, the scaling factor $\alpha$ is declared as a trainable parameter that is optimized using a gradient based approach.

For trained bit-width quantization we optimized the following loss function
\begin{equation}
	\label{eq:loss}
	L = L_{CE} + \lambda_w \cdot B_w + \lambda_a \cdot B_a
\end{equation}
where $L_{CE}$ is the cross-entropy loss, $\lambda_w$, $\lambda_a$ are hyperparameters and $B_w$, $B_a$ are the average weight and activation bit-width of the model respectively. For our experiments, the hyperparameters were selected as $\lambda_w = 0.04$ and $\lambda_a = 0.04$. Note that in brevitas, trained bit-width quantization is limited to bit-widths larger or equal to 2. However, brevitas is currently under active development and future versions may allow 1 bit activations and weights for trained bit-width quantization.

\section{Experimental setup}
\label{sec:experimental_setup}

\subsection{Dataset}
We use the first version of the Google speech commands dataset \cite{Warden2018}. It consists of 65.000 1-second long audio files sampled with 16 bit at 16 kHz sampling frequency. Every audio file contains one utterance of an English word spoken by one person. The words are grouped into 30 different classes. We follow the procedure of \cite{Tang2017} and use the following 10 classes ”Yes”, ”No”, ”Up”, ”Down”, ”Left”, ”Right”, ”On”, ”Off”, ”Stop” and ”Go” from the dataset. Likewise, we also include an "unknown" class which is a blend of randomly selected samples from the remaining 20 classes. Furthermore, a "silence" class is added. The "silence" class is artificially generated and consists of 1-second audio files containing a random slice of audio from a randomly selected noise sample provided by the Google Speech commands dataset. We follow the procedure of \cite{Tang2017} and perform data augmentation by applying a random time shift and adding background noise to the raw audio waveforms.

\subsection{Feature extraction}
End-to-end KWS models presented in this paper do not need any hand-crafted feature extraction since they directly classify keywords from the raw audio waveforms. However, we will later compare end-to-end KWS models with models that use MFCCs as input. For models using MFCCs as input, the raw audio waveforms are first filtered with a low pass filter $f_{\mathrm{low}}=40 \mathrm{Hz}$ and a highpass filter $f_{\mathrm{high}}=4 \mathrm{kHz}$. We then extract between 10 and 30 MFCCs per 40ms frame with a stride length of 20ms. The number of MFCCs is varied in the experiments.

Before performing classification on raw audio waveforms, we select a window length of 25ms and a hop length of 10ms to split up the raw audio waveform into frames. Therefore, at a sampling frequency $f_s=16 \mathrm{kHz}$, the SincConv filter length is 400 and the hop length is 160. Before filtering with the SincConv, a Hamming window is applied to the frames.

%\subsection{Performance evaluation}
%Like the training dataset, the validation and test dataset include 10 keyword classes, one unknown class and one silence class. The keyword classes and the unknown class in the validation and test dataset are not augmented. However, the silence class is augmented with background noise. Since the unknown class and the silence class may change depending on the seed used to init the random functions for selecting the unknown samples and the background noise samples, we perform 3 separate runs per model. Per run we select a seed and then train the model on the training dataset while validating its performance on the validation dataset. At the end of training we test the model on the test dataset. Depending on the experiment we report either the maximum or the average test accuracy of the 3 runs. 

\section{Experiments}
\label{sec:experiments}

\subsection{KWS from raw audio waveforms using NAS}
We compare our end-to-end KWS models to models using MFCCs \cite{Peter2020}. We introduce a width multiplier $m$ that scales the number of channels of the model by a factor of $m=1, m=0.5$ or $m=2$. The width multiplier does not effect the SincConv layer at the input. For the SincConv layer, the number of filters was set to 40, 60, and 80. After a width-multiplier and the number of SincConv filters is selected, NAS is performed to select the layers in stage (iii) and (iv) of our NAS model (c.f. Table~\ref{model_structure}). A tradeoff between the model size and accuracy is established by varying the regularization parameter $\beta \in \{0, 1, 2, 4, 8, 16\}$ to obtain end-to-end KWS models of different sizes (for details on $\beta$ see \cite{Peter2020}). To assess the difference between SincConvs and 1D-Convs, we also performed NAS using models with 1D-Conv instead of SincConv. However, here we only select a width-multiplier of $m=1$.

\begin{figure}[t]
	\centerline{\includegraphics[width=\linewidth]{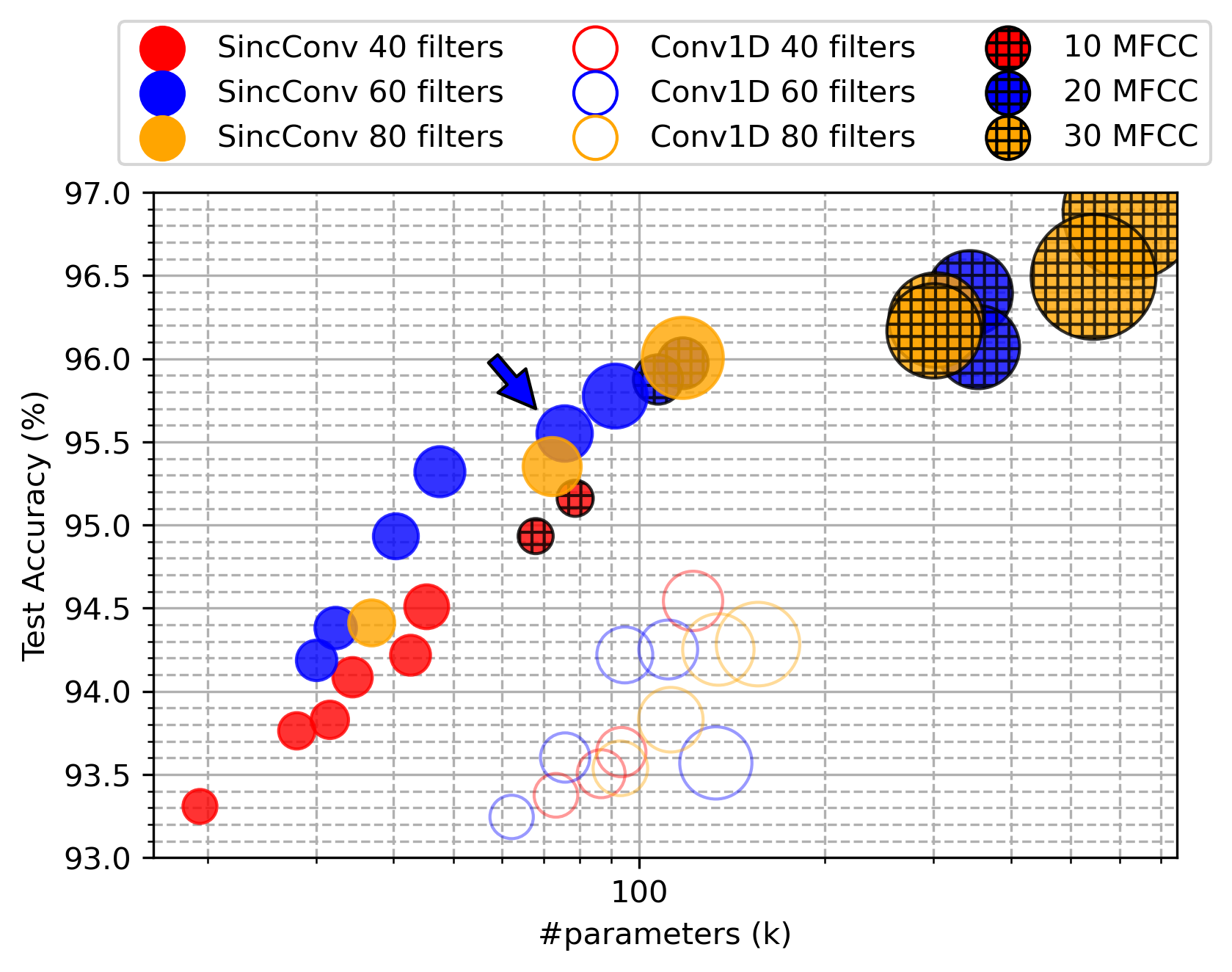}}
	\caption{Test accuracy vs number of parameters of KWS models obtained using NAS. Models on the Pareto frontier are emphasized. The number of operations corresponds to the circle area. The model marked with an arrow is quantized in Section~\ref{sec:quantization_exp}.}
	\label{fig:full_performance}
\end{figure}

Figure~\ref{fig:full_performance} shows the performance of SincConv and MFCC models. For better visibility, we only include models on the Pareto frontier. We also include 1D-Conv models although none of the models contributes to the Pareto frontier. The number of operations of a model corresponds to the circle area. We can observe that 1D-Conv models perform worse than SincConv models with regards to test accuracy, number of operations and number of parameters. This indicates that using parametrized sinc functions instead of fully learnable filter kernels provide a substantial benefit in the performance of end-to-end KWS models. We can also observe that SincConv models need less parameters than MFCC models to achieve the same test accuracy. However, the number of operations is slightly larger in SincConv models.

Using NAS only, we were able to obtain efficient models. We highlighted one model from Figure~\ref{fig:full_performance} (marked with a blue arrow) that will be quantized in Section~\ref{sec:quantization_exp}. This model achieves an accuracy of 95.55\% using only 75.7k parameters and 13.6M operations.

\subsection{Weight and activation quantization}
\label{sec:quantization_exp}
\begin{figure}[t]
	\centerline{\includegraphics[width=1.06\linewidth]{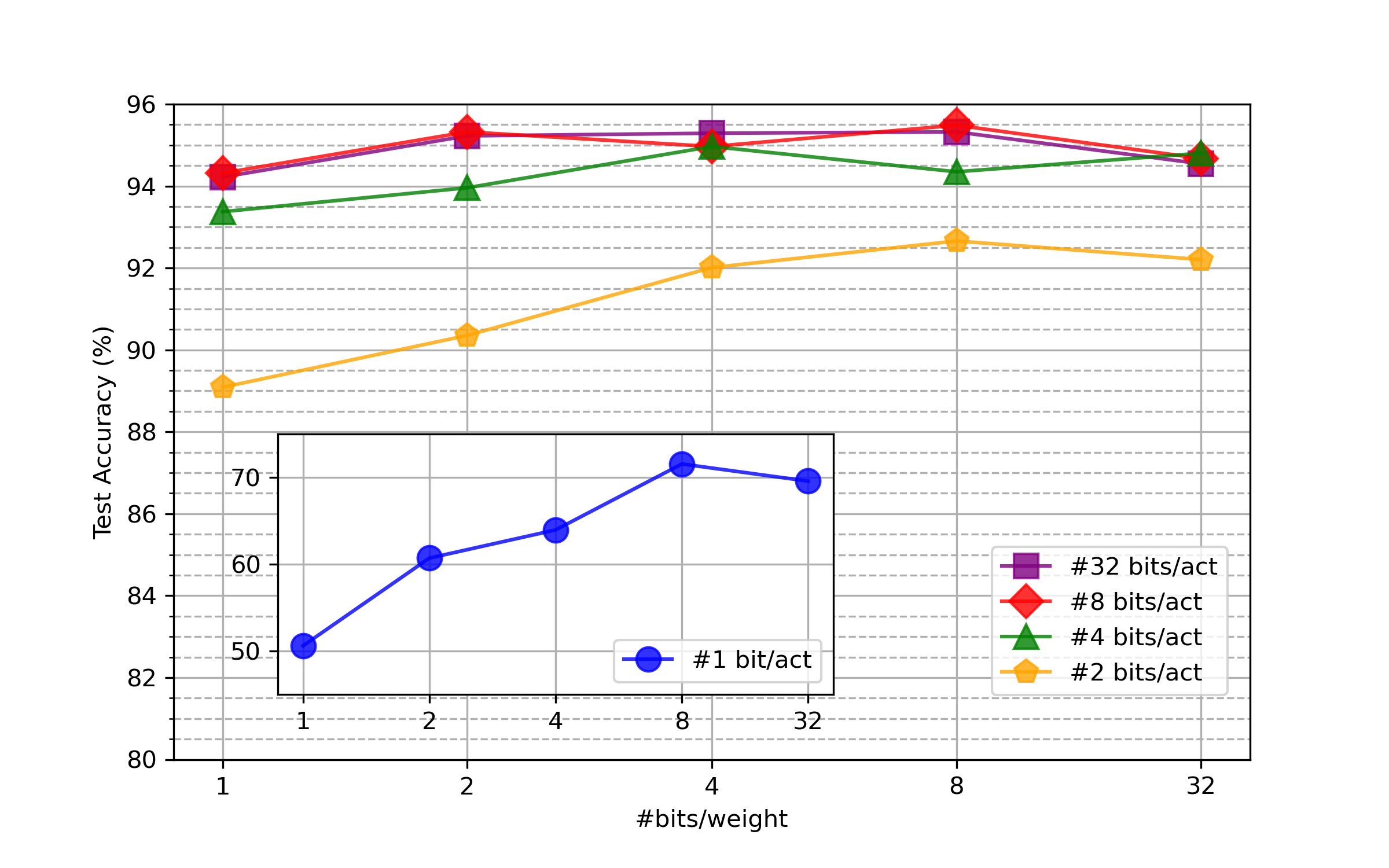}}
	\caption{Test accuracy vs weight bit-width vs activation bit-width of an end-to-end KWS model using SincConvs. The model was trained from scratch using quantization-aware training and fixed bit-widths.}
	\label{fig:fixed_quantization}
\end{figure}

After NAS is performed and a suitable full-precision end-to-end KWS model is found, we quantize the weights and activations to further reduce the memory footprint of the model. We select the model marked with a blue arrow from Figure~\ref{fig:full_performance}. We first perform fixed bit-width quantization for the weights and activations. The results are visualized in Figure~\ref{fig:fixed_quantization}.

We observe that the most notable impact on performance is encountered when the activation bit-width is reduced to 2 or even 1 bit. When the activation bit-width is 4 bits, a slight performance impact is observed. However, when using 8 bit activations, the performance is similar to the full precision model irrespective of the weight bit-width. 

\begin{figure}[t]
	\centerline{\includegraphics[width=\linewidth]{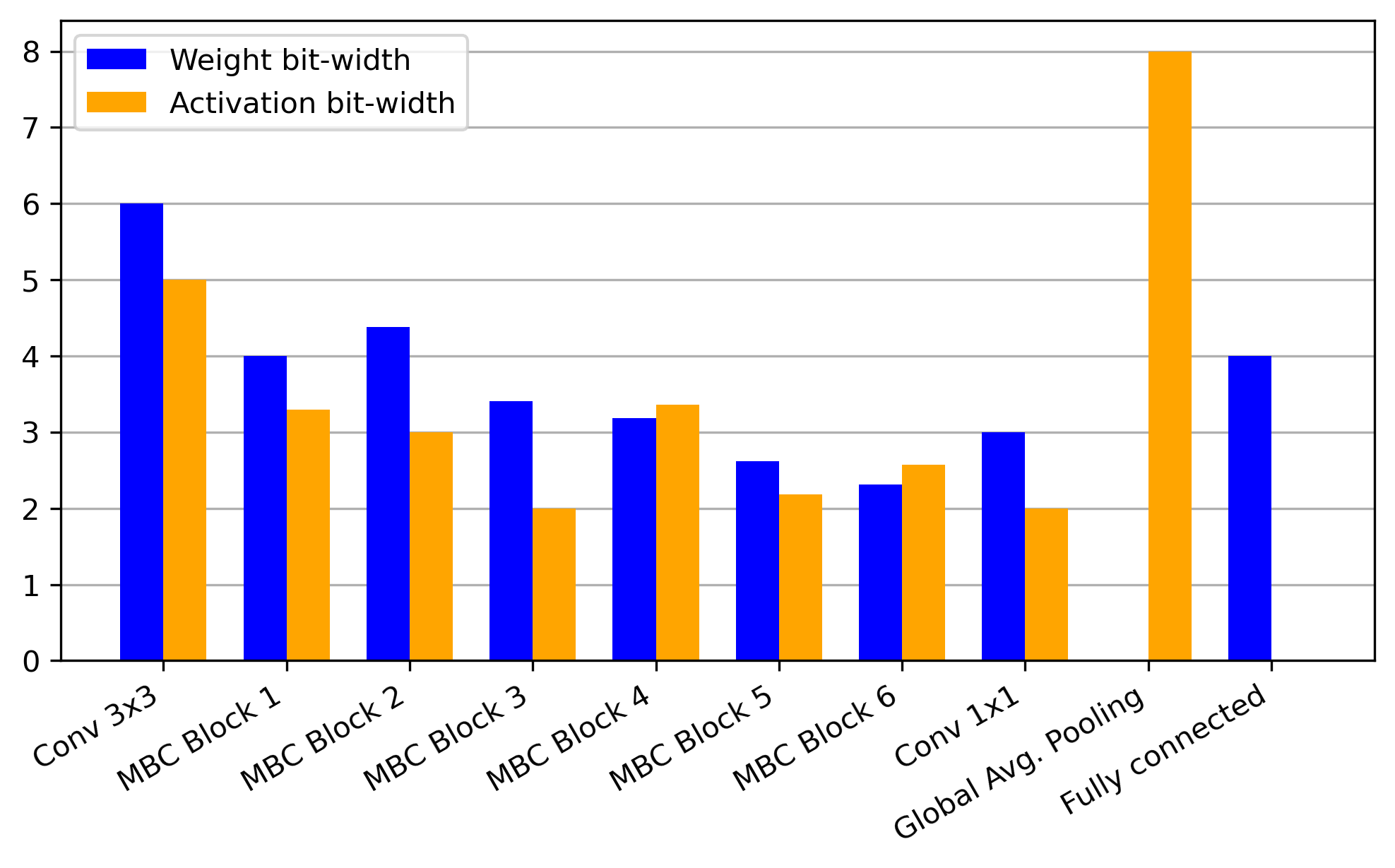}}
	\caption{Weight and activation bit-widths of an end-to-end KWS model using SincConvs. The model was trained from scratch using quantization-aware training and trained bit-widths.}
	\label{fig:trained_quantization}
\end{figure}

We also perform trained bit-width quantization on the same model. However, now the weight and activation bit-width are optimized together with model parameters using backpropagation. The weight and activation bit-widths per layer after training are visualized in Figure~\ref{fig:trained_quantization}. For MBC blocks, we report the average per-weight and per-activation bit width over the three convolutions.

The trained bit-width model visualized in Figure~\ref{fig:trained_quantization} needs on average 2.91 bits for quantized activations and 2.51 bits for quantized weights and achieves a test accuracy of 93.76\%. Compared to a fixed bit-width model with 2 bit activations and 2 bit weights with a test accuracy of 90.35\%, the trained bit-width model outperforms the fixed bit-width model by 3.41\% while using only slightly more bits on average. However, hardware implementation of trained bit-width quantization is more difficult.

We also observe that layers at the input and the output need more bits than intermediate layers. This result is in line with the vast literature where it is common practice to leave the input and output layers at full precision.

\section{Conclusions}
\label{sec:conclusions}
Resource-efficient DNNs are the key components in modern keyword spotting (KWS) systems. We used neural architecture search (NAS) to obtain efficient end-to-end convolutional neural networks (CNNs) for KWS. Our end-to-end KWS models utilize a SincConv at the input layer to perform classification on raw audio waveforms. To make our results comparable, we performed NAS on the Google speech commands dataset. We compared our end-to-end KWS models to mel-frequency cepstral coefficient (MFCC) based systems. We also compared two weight and activation quantization methods that help to further reduce the memory footprint. By establishing a tradeoff between the model accuracy and the model size, we show that trained bit-width quantization can be used to obtain more competitive models than simply using fixed bit-width quantization. 

\pagebreak
\bibliographystyle{IEEEtran}

\bibliography{mybib}

% \begin{thebibliography}{9}
% \bibitem[1]{Davis80-COP}
%   S.\ B.\ Davis and P.\ Mermelstein,
%   ``Comparison of parametric representation for monosyllabic word recognition in continuously spoken sentences,''
%   \textit{IEEE Transactions on Acoustics, Speech and Signal Processing}, vol.~28, no.~4, pp.~357--366, 1980.
% \bibitem[2]{Rabiner89-ATO}
%   L.\ R.\ Rabiner,
%   ``A tutorial on hidden Markov models and selected applications in speech recognition,''
%   \textit{Proceedings of the IEEE}, vol.~77, no.~2, pp.~257-286, 1989.
% \bibitem[3]{Hastie09-TEO}
%   T.\ Hastie, R.\ Tibshirani, and J.\ Friedman,
%   \textit{The Elements of Statistical Learning -- Data Mining, Inference, and Prediction}.
%   New York: Springer, 2009.
% \bibitem[4]{YourName17-XXX}
%   F.\ Lastname1, F.\ Lastname2, and F.\ Lastname3,
%   ``Title of your INTERSPEECH 2021 publication,''
%   in \textit{Interspeech 2021 -- 20\textsuperscript{th} Annual Conference of the International Speech Communication Association, September 15-19, Graz, Austria, Proceedings, Proceedings}, 2020, pp.~100--104.
% \end{thebibliography}

\end{document}